\documentclass{template/llncs}

\usepackage{times}

\usepackage{comment}

\usepackage{graphicx}

\usepackage[english]{babel}

\usepackage{array}

\excludecomment{quote}

\usepackage{hyperref}

\def\plaintitle{Estimating Visual Comfort in Stereoscopic Displays Using Electroencephalography: A Proof-of-Concept}
\def\plainauthor{J\'er\'emy Frey, Aur\'elien Appriou, Fabien and Martin Hachet}
\def\plainkeywords{Stereoscopy; Comfort; EEG; Adaptive system; Evaluation}

\hypersetup{
pdftitle={\plaintitle},
pdfauthor={\plainauthor},
pdfkeywords={\plainkeywords}
}

\begin{document}

\title{\plaintitle}

\author{J\'er\'emy Frey \inst{1,2,3} \and Aur\'elien Appriou \inst{3,1} \and Fabien Lotte \inst{3,2} \and Martin Hachet \inst{3,2}}


\institute{Univ. Bordeaux, LaBRI, UMR 5800, F-33400 Talence, France \and CNRS, LaBRI, UMR 5800, F-33400 Talence, France \and INRIA, F-33400 Talence, France \\
\email{first\_name.last\_name@inria.fr}}

\maketitle              

\begin{abstract}

With stereoscopic displays, a depth sensation that is too strong could impede visual comfort and result in fatigue or pain. Electroencephalography (EEG) is a technology which records brain activity. We used it to develop a novel brain-computer interface that monitors users' states in order to reduce visual strain. We present the first proof-of-concept system that discriminates comfortable conditions from uncomfortable ones during stereoscopic vision using EEG. It reacts within 1s to depth variations, achieving 63\% accuracy on average and 74\% when 7 consecutive variations are measured. 
This study could lead to adaptive systems that automatically suit stereoscopic displays to users and viewing conditions. 

\keywords{\plainkeywords}
\end{abstract}

\section{Introduction}\label{introduction}

Stereoscopic displays have been developed and used for years in computer science, for example to improve data visualization 
or to better manipulate virtual objects \cite{Drossis2013}. However, only recently did they begin to reach users beyond experts, at home, with ``3D'' movies or game consoles.
Yet, whenever devices use shutter or polarized glasses, parallax barrier or head-mounted displays (as with the Oculus Rift) to produce pairs of images, visual discomfort could occur when the stereoscopic effect is too strong. 
Some viewers could even suffer pain \cite{Lambooij2009}.

In order to mitigate those symptoms and adapt the viewing experience to each user, we propose an innovative method which can discriminate uncomfortable situations from comfortable ones. It reacts quickly (within 1s), without calling upon users, so it does not disrupt the viewing. Our solution is versatile because all stereoscopic displays use the same mechanism to give the illusion of depth. They send a different image to the left and right eyes. As with natural vision, the visual fields of our eyes overlap and the difference between the two images helps our brain to estimate objects' distance.

To facilitate images merge, observers rely on two mechanisms. First, they need to maintain the point of interest at the same place on both their retinas. This is why the closer an object gets, the more eyeballs rotate inward. This is called ``vergence'', and it also happens with stereoscopic displays. Second, in a way similar to how camera lenses operate, crystalline lenses need to focus light beams. They deform accordingly to objects' position in order to obtain a clear picture. This other physiological phenomenon is called ``accommodation'' and is \emph{not} replicated with stereoscopic displays.


In a natural environment, vergence and accommodation are locked to objects' positions and occur altogether. But since the focal plane in stereoscopic displays is fixed, accommodation will not change. 
The discrepancy between vergence and accommodation is called the ``vergence-accommodation conflict'' (VAC).
When it is too strong or lasts too long, visual discomfort occurs \cite{Lambooij2009}. VAC is one of the major causes of the symptoms associated with visual discomfort and visual fatigue in stereoscopic displays \cite{Lambooij2009}. Guidelines exist to limit the VAC and prevent such negative effects. In particular, Shibata et al. \cite{Shibata2011} established a ``zone of comfort'' using questionnaires, a zone within which the apparent depth of objects should remain to avoid discomfort.
It takes into account the distance between viewers and displays. Unfortunately, screen settings, viewing angle or individual differences \cite{Lambooij2009} make it hard to generalize such recommendations and use them as is.


Complementary to qualitative questionnaires, as used in \cite{Shibata2011}, brain activity recordings enable the monitoring of users' states \cite{Frey2014a,Zander2011}. One of the main advantages of such a technology for the evaluation of human-computer interaction (HCI) comes from the real-time insights that it could give. 
Electroencephalography (EEG) is among the cheapest and most lightweight devices that sense brain signals. Even though EEG has been used to investigate visual fatigue induced by stereoscopic display \cite{Li2008a,Cho2012,Bang2014}, those studies only compared flat images with stereoscopy. They do not control for objects virtual positions, hence they cannot account for different comfort conditions. Furthermore, most of the EEG studies related to stereoscopic display and comfort analyzed stimuli which last several minutes. Such protocols could not lead to adaptive systems that react quickly; they focus more on the overall fatigue induced by a prolonged exposition to stereoscopy rather than discomfort. 
In \cite{Frey2014}, we conducted a preliminary investigation that compared short appearances of virtual objects. We showed that the average brain activity induced by stereoscopic displays was different whether objects were presented within the zone of comfort or not. 
Furthermore, we set the basis of a protocol which uses short presentation time (few seconds). However, the study involved only 3 participants and only investigated the averaged EEG signal, with no classification of the events.

Inspired by these works, we designed and tested a system that classifies EEG data to measure visual comfort. This type of system is a \emph{passive} brain-computer interface (BCI, \cite{Zander2011}). 
Our main contribution is to prove the feasibility of an EEG system that could estimate in near real-time (1s delay) the visual comfort viewers are experiencing as they watch stereoscopic displays. It could be adapted to real-case scenarios by controlling the discrepancy between left and right images depending on the output of the classifier. Then it could be employed in different settings to improve HCI by easing users' comfort, for example when they manipulate 3D contents during prolonged periods of time -- remote design, video games, and so on -- or when people are watching 3D movies -- especially when there are many rapid depth variations, as seen in action sequences.

\section{Experiment}\label{experiment}

\subsection{Overview}\label{overview}

\begin{figure}[htbp]
\centering
\includegraphics[width=0.75\textwidth]{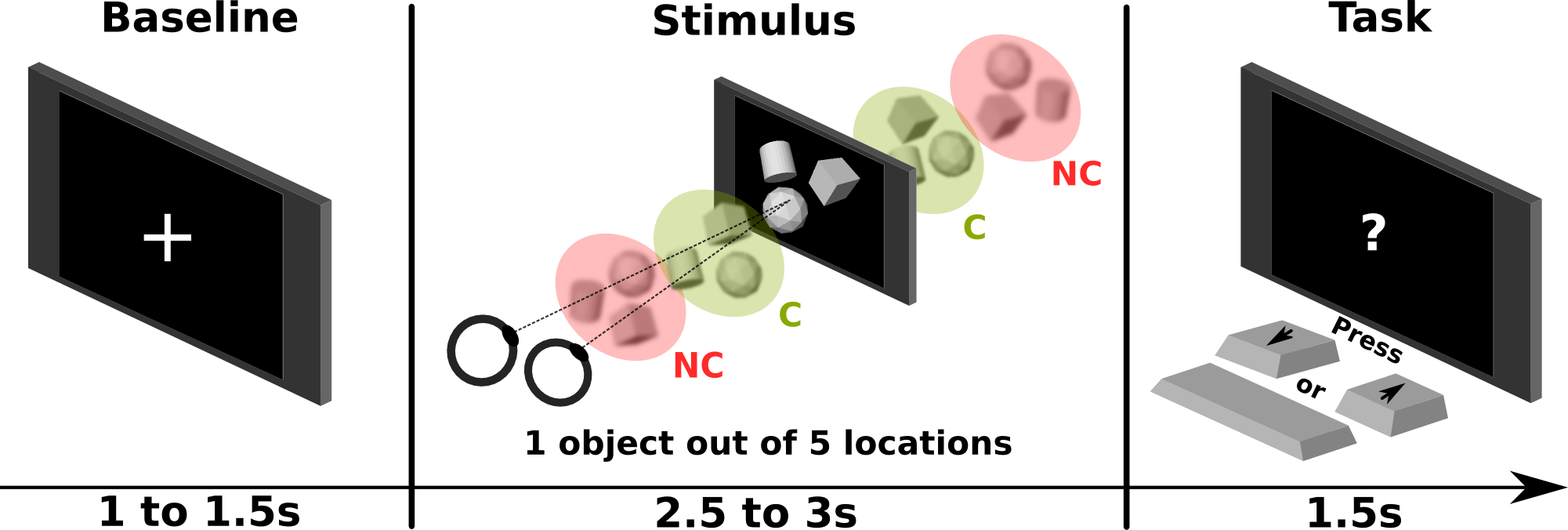}
\caption{One trial: cross (baseline), object at random depth, task.\label{fig:trial}}
\end{figure}

We studied the appearance of virtual objects. They were presented to participants at different apparent depths for a few seconds (see Figure \ref{fig:trial}). We created two conditions: objects appeared either at a comfortable position (``C'' condition) or at an uncomfortable position (``NC'' condition). We displayed simple grey objects over a black background. Three kinds of primitives were employed: cube, cylinder and icosphere.
Objects' orientations were randomized along the three axes to create various stimuli.
Rotations were controlled so as the faces of cubes and cylinders could not be orthogonal to the camera plane, thus preventing the appearance of artificial 2D shapes. The resulting 3D scenes were kept simple enough to ensure that there were no distracting elements and that no variables besides the VAC were manipulated. We deprived the depth cues to control for VAC. For example casting shadows would have helped to differentiate close objects from far objects without the need of binocular fusion \cite{Mikkola2010}.

We defined ranges inside and outside the zone of comfort according to \cite{Shibata2011}. Related to the location of participants sitting 1m away from the display, in ``C'' condition virtual objects were positioned within [0.75m; 0.85m] (comfortable close) or within [1.3m; 1.6m] (comfortable far). In ``NC'' conditions, ranges were [0.35m; 0.45m] (uncomfortable close) or [4m; 6m] (uncomfortable far). During one-third of the trials, objects appeared ``flat'' (no stereoscopic effect, 1m apparent depth, as far as the screen).

In order to assess their capacity to situate virtual objects in space and to maintain their vigilance high during the whole experiment, participants had to perform a task. When a question mark was shown on screen, ``down'', ``space'' or ``up'' keys were pressed to indicate whether objects appeared ``in front of'', ``as far as'' or ``behind'' the screen. With both hands on the keyboard, choosing those keys to answer ensured that participants' gaze was not leaving the screen and that participants movements would not pollute EEG signals.

A trial started with a neutral stimulus, a 2D cross appearing on-screen for a duration comprised between 1 and 1.5s (Figure \ref{fig:trial}). Then the virtual object appeared for 2.5 to 3s. Finally, a question mark appeared for 1.5s, a period during which participants had to perform the task. After that a new trial began. The first two time intervals, that randomly varied by 0.5s, prevented participants to anticipate objects appearance and the moment they had to respond to the task. On average a trial took 5.5s. There were 160 trials per C and NC conditions, randomly distributed. Trials were also equally split across 4 sub-sessions to let participants rest during the investigation and avoid a too tedious experiment.

\subsection{Apparatus}\label{apparatus}

Stereoscopic images were shown in 1080p resolution on a 65 inches Panasonic TX-P65VT20E active display -- participants wore shuttered glasses. 
No matter the apparent depths of displayed objects, their sizes on screen remained identical. 
In combination with a diffuse illumination of the scene, this made it impossible to discriminate conditions without stereoscopy. The interpupillary distance used to compose stereoscopic images was set at 6cm, an average value across population \cite{Dodgson2004}. EEG signals were acquired at a 512Hz sampling rate with 2 g.tec g.USBamp amplifiers. 
We used 4 electrodes to record electrooculographic (EOG) activity and 28 to record EEG. In the international 10-20 system, EOG electrodes were placed at LO1, LO2, IO1 and FP1 sites; EEG electrodes were placed at AF3, AF4, F7, F3, Fz, F4, F8, FC5, FC1, FC2, FC6, C3, Cz, C4, CP5, CP1, CP2, CP6, P7, P3, Pz, P4, P8, PO3, PO4, O1, Oz and O2 sites. 

\subsection{Participants}\label{participants}

12 participants took part in the experiment: 5 females, 7 males, mean age 22.33 (SD=1.15). They reported little use of stereoscopic displays: 1.91 (SD=0.54) on a 5-point Likert scale (1: never; 2, 3, 4, 5: several times a year/month/week/day respectively). If applicable, participants wore their optical corrections -- there was enough space beneath the shutter glasses for regular glasses not to disrupt user experience. We made sure that no participant suffered from stereo blindness by using a TNO test \cite{Momeni-Moghadam2012}.

\subsection{Measures}\label{measures}

Beside EEG measures, task scores were computed from participants' assessment of objects' virtual position in space -- whether they appeared ``in front of'', ``as far as'' or ``behind'' the screen. During the 1.5s time window when question marks appeared, the first key pressed, if any, was taken into account. A correct answer resulted in 1 point, an incorrect in -1 point and none in 0 point. Final scores were normalized from [-480;480] to [-1;1] intervals.

A questionnaire inquiring the symptoms associated with the different apparent depths preceded first trials and followed each sub-session. There were 2 items, one asking about participants vision clarity and the other about eyes tiredness. The corresponding 5-point Likert scales were adapted from \cite{Shibata2011}, ``1'' representing no negative symptoms and ``5'' severe symptoms. We measured respectively how well participants saw the stereoscopic images and how comfortable they felt, averaging in total 10 answers for each C/NC conditions.

\subsection{Procedure}\label{procedure}

The experiment occurred in a quiet environment, with a dimmed ambient light and was approximately 90 minutes long. It comprised the following steps: 

1: Participants were seated 1m away from the stereoscopic screen (distance from their eyes) and filled an informed consent form and a demographic questionnaire.

2: Participants' stereoscopic vision was assessed with a TNO test.

3: An EEG cap was installed onto participants' heads.

4: The ``symptoms'' questionnaire was given orally, experimenter manually triggering objects appearances and waiting for participants' answers during this phase. There was 1 object per virtual depth range (C close/far, NC close/far) and 2 flat objects; making 6 randomized objects per questionnaire.

5: A training session occurred. During this session participants had the opportunity to get familiar with the trials and with the task. 

6: The 4 sub-sessions, described previously, occurred. When a sub-session ended, participants were given again the questionnaire of step 4 before they could rest, drink and eat. Once they felt ready, we pursued with the next sub-session.

\section{Analyses}\label{analyses}

\label{sec:analyses}



Data gathered from the 4 sub-sessions were concatenated. We applied a 0.5Hz high-pass filter to correct DC drift and a 25Hz low-pass filter to remove from our study signal frequencies that were more likely to be polluted by muscle activity. We extracted the 320 epochs -- ``slices'' of EEG -- around C and NC stimuli onsets, from -1s to +2.5s. Due to the substantial amount of data (3840 trials in total), we chose automated methods to clean the signals. The EEGLAB (\url{http://sccn.ucsd.edu/eeglab/}) function pop\_autorej removed epochs that contained muscular artifacts ($\approx$ 10\%). Following the results from \cite{Ghaderi2013}, EOG activity was suppressed using the ADJUST toolbox 1.1. After an Infomax independent component analysis (ICA), we removed from the original signal components that ADJUST labeled as eye blinks or eye movements (vertical and horizontal).

\begin{figure}[htbp]
\centering
\includegraphics[width=0.75\textwidth]{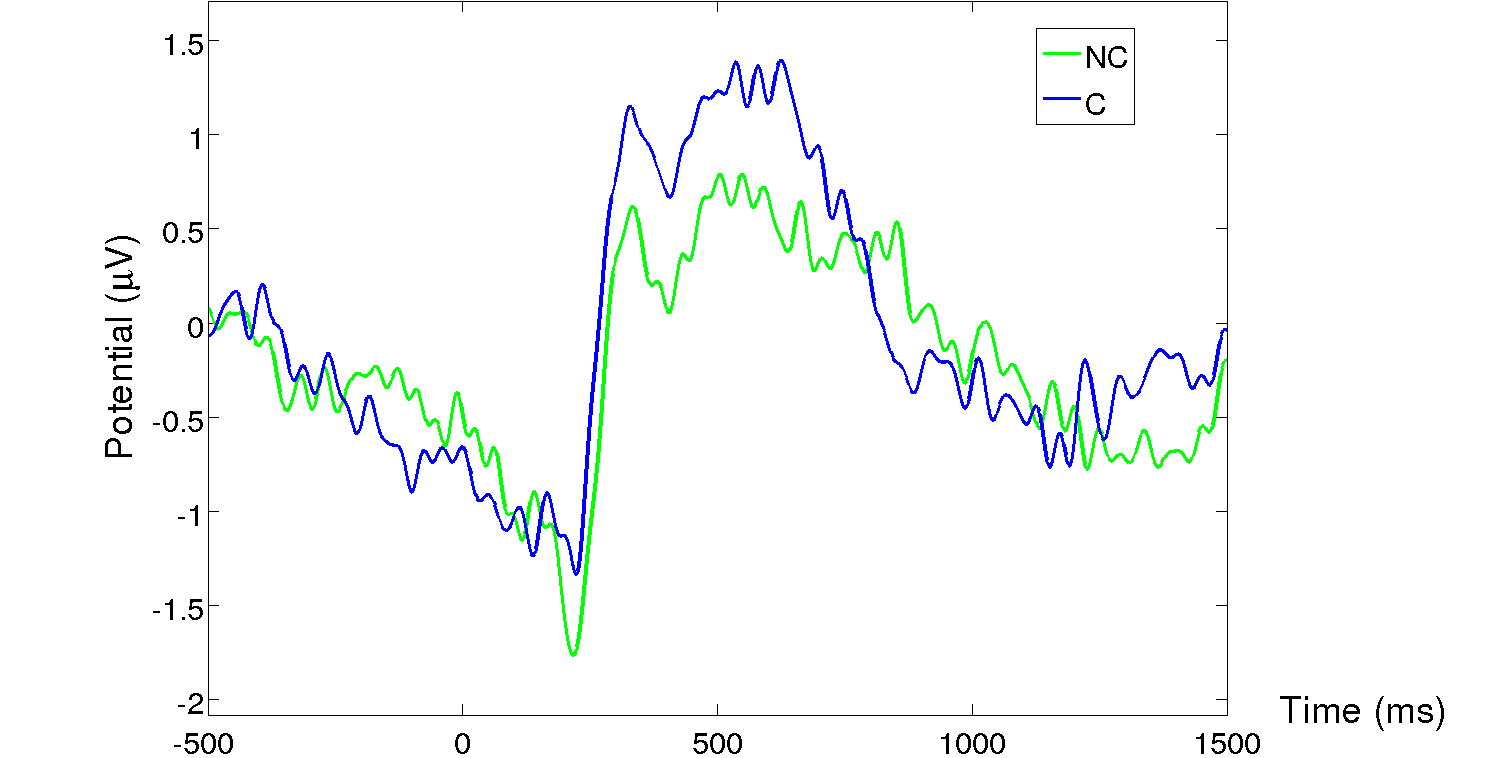}
\caption{Average ERP across 28 EEG electrodes and 12 participants. Blue: comfort condition; green: no-comfort condition ($\approx$ 160 trials each). The stereoscopic object appears at t=0ms.\label{fig:erp}}
\end{figure}

An event-related potential (ERP) corresponds to one or more ``peaks'' in EEG recordings, associated with an event -- in our case the appearance of stereoscopic images. Averaged ERPs across participants indicated that ERPs had a higher positive peak in C (see Figure \ref{fig:erp}). Note that there were also some differences in EEG oscillations (not reported here due to space limitations), although less salient. Thus, in this study we only used ERP, as using EEG oscillations did not lead to any substantial improvement.

For classification, we first split the EEG dataset of each participant in two. The first half of the trials was used as a training set, to calibrate the classifier. The second half was used as a testing set, to test the classifier performances on unseen data, which simulates a real-case application. 
Feature extraction relied on the spatial filter from \cite{Hoffmann2006}, which was specifically designed for ERPs classification, and reduced signals dimension from 28 EEG channels to 5 ``virtual'' channels whose signal is more discriminant between conditions. 
We selected a time window of 1s.
In order to reduce the number of features, we decimated the signal by a factor 16. As a result, there was 160 features by epoch (5 channels $\times$ 512Hz $\times$ 1s / 16). We used shrinkage Linear Discriminant Analysis as a classifier, as recommended in \cite{Blankertz2011} for ERP classification. 


Although we used 1s time windows as a basis for our analyses, we tested longer stimuli by clustering trials with Monte Carlo simulations.
The principle is as follows: studying 3 presentations, we cluster 3 similar trials drawn from the testing set (e.g., ``no-comfort'', 3xNC). Then we look at individual classification results from the system (e.g., NC-NC-C) and keep the label which has the majority -- in this case NC, the resulting classification is correct for this cluster. Had the classifier labelled trials as ``C-C-C'',
the cluster would have been erroneously labeled as ``C''. Different combinations of trials were drawn from the testing set to compute the scores for n=3,5,7. 

\section{Results}\label{results}


We used a Wilcoxon Signed-rank test to compare task scores between C and NC conditions. 
There was no significant effect (p = 0.78).
There was a significant effect of the C/NC conditions on both symptoms items (p \textless{} 0.01). Participants reported more eye comfort (means: 2.41 \emph{vs} 3.46) and more vision clarity (means: 2.10 \emph{vs} 3.13) in C than in NC. We were able to predict with an average classification accuracy of 63.30\% (SD=7.64) the visual comfort experienced by viewers (see Table \ref{tab:class}).



\begin{table}[htbp]
\centering
\caption{Classifier accuracy (in percentage) for each user. Mean: 63.30\%, SD: 7.64.}
\begin{tabular}{l|p{0.5cm} p{0.5cm} p{0.5cm} p{0.5cm} p{0.5cm} p{0.5cm} p{0.5cm} p{0.5cm} p{0.5cm} p{0.5cm} p{0.5cm} p{0.5cm}}
  \textbf{User} & 1 & 2 & 3 & 4 & 5 & 6 & 7 & 8 & 9 & 10 & 11 & 12 \\
  \hline
  \textbf{Accuracy} & 54.2 & 59.2 & 58.2 & 70.3 & 60.5 & 64.2 & 62.9 & 76.1 & 72.5 & 71.5 & 53.2 & 56.7 \\  
\end{tabular}
\label{tab:class}
\end{table}

With Monte Carlo simulations, we investigated how the system would perform with the appearance of several images from the same condition. Classifier accuracy reached 68.91\% (SD=10.32) over 3 trials. Over 5 trials the classification reached 90\% for some users, resulting in a 71.83\% average (SD=12.28). With n=7, one-third of the participants reached 90\% or more (74.08\% on average, SD=13.39).

The pipeline that we presented in Section \ref{sec:analyses} would be difficult to apply in real-life scenarios -- online analyses prevent the use of advanced signal processing, such as ICA for artifact removal, because it requires heavy computations. 
Fortunately, artifacts had little incidence on the performance of our system.
We obtained similar results without artifacts denoising when we simulated online analyses, with an accuracy of 62.40\% (SD=4.80).

\section{Discussion \& Conclusion} \label{discussion}\label{conclusion}

During short exposures to images, participant reported worse vision clarity and less visual comfort in NC condition, thereby validating a clear distinction between the two zones of comfort of our protocol. Participants performed equally well in both conditions during the task, suggesting that even if severe, a VAC does not alter their ability to make rough estimations of virtual depths. In this context, it also highlights the limits of behavioral methods for measuring participants' comfort. A neuroimaging technique, on the other hand, did manage to discriminate two comfort conditions.

EEG signals reflected the disparities in visual comfort. 
It was possible to build a classifier that achieved an accuracy greater than 63\%, with several participants exceeding 70\%. The system scored above chance level in all our analyses (p \textless{} 0.01) \cite{Muller-Putz2008}. 

This score of 63\% accuracy, while not as high as some other established BCI systems, may be already sufficient to improve users' comfort. Indeed, on-the-fly correction of uncomfortable images can be seen as error correction. In such settings, detection rates from 65\% are acceptable to improve interactions \cite{Vi2012}. These findings depend on the nature of the task; this is why we proposed a mechanism to increase the performance of the classifier by taking into account more than one object appearance. The system score improved by 6 points when we clustered trials by 3. During our simulations, the accuracy reached 90\% for some users with 5 trials, and for one-third of the participants over 7 trials. 
Therefore, this tool can estimate how many presentations are needed to reach a specific accuracy and suit the desired application.
Interestingly enough, when using less EEG channels (only 14) we obtained similar classification accuracies -- results not reported here due to space limitations -- which suggests that our approach might be used with entry-level devices such as the Emotiv EPOC (\url{https://emotiv.com/}) or the OpenBCI system (\url{http://www.openbci.com/}).




We described an innovative system that can distinguish uncomfortable stereoscopic viewing conditions from comfortable ones by relying on EEG signals. We controlled the experimental conditions with questionnaires, founding significant differences in visual comfort between short exposures of images. Visual \emph{comfort} was assessed, whereas existing studies focused on visual \emph{fatigue} -- a component that appears on the long term and that we propose to prevent beforehand.

Using short time windows (features were extracted over 1s), we set the basis of a tool capable of monitoring user experience with stereoscopic displays in near real-time.
Our offline analysis
demonstrated the feasibility of such a method with clean EEG signals. We obtained a similar classification accuracy without computationally demanding artifacts filtering, demonstrating also that the work presented here could perfectly be applied online. Ongoing evaluations include considering additional EEG features such as EEG band power to further boost performances.
The code that generated the images and the scripts that we used to process signals will be released in order to ease replication, in the hope that this combination of EEG and HCI evaluation will benefit end users.


Such a passive BCI can adapt the parameters to users' state (e.g., mental fatigue is likely to relate to visual fatigue) throughout the viewing. Moreover, a passive BCI does not disrupt work or the narrative of the stereoscopic environment.

A passive stereoscopic comfort detector could potentially be useful for multiple applications, as a tool to: 1) objectively compare (possibly offline) different stereoscopic displays, 2) dynamically enhance stereoscopic effects, by increasing discrepancy without causing discomfort, 3) quickly calibrate stereoscopic displays, 4) dynamically adapt discrepancy to avoid discomfort (e.g., during 3D movies) or voluntarily cause discomfort (e.g., for basic science studies about perception), among many others. 

We documented a novel solution to a famous issue -- i.e., estimating stereoscopic discomfort -- thus increasing fundamental knowledge. Besides 3D scenes control, by giving access in real-time to users' inner states, EEG will help to modulate more closely the viewing experience according to the effect one wants to achieve.

\bibliographystyle{template/splncs03}
\bibliography{stereo2_short.bib}

\end{document}